\pdfoutput=1
\documentclass{moriond}

\def\Journal#1#2#3#4{{#1} {\bf #2}, #3 (#4)}

\def\PRL{\em Phys. Rev. Lett.}
\def\PRD{{\em Phys. Rev.} D}

\def\be{\begin{equation}}
\def\ee{\end{equation}}
\def\bea{\begin{eqnarray}}
\def\eea{\end{eqnarray}}

\usepackage{ifthen}
\newboolean{uprightparticles}
\setboolean{uprightparticles}{false} %Set true for upright particle symbols
\usepackage{xspace} 
\usepackage{upgreek}

\def\Pnu         {\ensuremath{\nu}\xspace}
\def\neub       {{\ensuremath{\overline{\Pnu}}}\xspace}
\def\neutb      {{\ensuremath{\neub_\tau}}\xspace}

\def\PB      {\ensuremath{B}\xspace}
\def\Bbar    {{\ensuremath{\kern 0.18em\overline{\kern -0.18em \PB}{}}}\xspace}
\def\PD      {\ensuremath{D}\xspace}                 
\def\Dbar    {{\kern 0.2em\overline{\kern -0.2em \PD}{}}\xspace}
\def\Dzb     {{\ensuremath{\Dbar{}^0}}\xspace}
\newcommand{\etal}{\mbox{\itshape et al.}\xspace}

\newcommand{\Photo}{}

\begin{document}
\vspace*{4cm}
\title{MEASUREMENT OF $\mathcal{R}(D^*)$ WITH THREE-PRONG $\tau$ DECAYS AT LHCb}

\author{ F. BETTI on behalf of the LHCb Collaboration }

\address{Universit\`a di Bologna, Dipartimento di Fisica e Astronomia,\\
Istituto Nazionale di Fisica Nucleare - Sezione di Bologna,\\
viale Berti Pichat 6/2, Bologna (40127), Italy}

\maketitle\abstracts{
The observable $\mathcal{R} ( D^{(*)} ) = \mathcal{B}\left( B^{0}\to D^{(*)-} \tau^{+} \nu_{\tau} \right) / \mathcal{B}\left( B^{0}\to D^{(*)-} \mu^{+} \nu_{\mu} \right)$ is a probe for Lepton Universality violation, so it is sensitive to New Physics processes.
The current combination of the measurements of $\mathcal{R} ( D^{(*)} )$ differs from Standard Model predictions with a $4\sigma$ significance.
A measurement of $\mathcal{R} ( D^* )$ using three-prong $\tau$ decays is currently ongoing at LHCb.
The statistical precision of this analysis is 6.7\%, i.e. the smallest statistical uncertainty for a single measurement of this observable.
Therefore this measurement will be important to confirm or disprove the current discrepancy from the theoretical expectations.
}

\section{Introduction}

In the Standard Model (SM) of particle physics the electroweak couplings of the gauge bosons to the leptons are independent of their flavour, a property known as lepton universality (LU), so the observation of LU violation would be a clear signal of physics processes beyond the SM.

The branching fractions ratio:
\begin{equation}
	\mathcal{R} ( D^{(*)} ) = \frac{ \mathcal{B}\left( B^{0}\to D^{(*)-} \tau^{+} \nu_{\tau} \right) }{ \mathcal{B}\left( B^{0}\to D^{(*)-} \mu^{+} \nu_{\mu} \right) }
\label{eq:RDstar}
\end{equation}
represents a sensitive probe for LU violation.

The combination of the measurements of $\mathcal{R} ( D^{(*)} )$ already performed by BaBar~\cite{babar2013}, Belle~\cite{belle2015,belle2016_1,belle2016_2} and LHCb~\cite{lhcb2015} shows a discrepancy of about $4 \sigma$ with respect to the values of $\mathcal{R}( D^{(*)} )$ calculated within the SM~\cite{fajfer2012}.% (see Figure~\ref{fig:HFAG}).

This document presents the analysis strategy and the perspectives of the measurement of $\mathcal{R} ( D^* )$, using three-prong $\tau$ decays, which is currently performed at LHCb with data collected during 2011 and 2012 at a centre-of-mass energy of 7 and 8 TeV, corresponding to an integrated luminosity of 3 $\textrm{fb}^{-1}$.

\section{Analysis Strategy}

The signal chosen for the analysis is $B^0 \to D^{*-} \tau^+ \nu_\tau$, where the $D^{*-}$ is reconstructed through the $D^{*-} \to \Dzb( \to K^+ \pi^-) \pi^-$ decay chain, while the $\tau$ lepton is reconstructed through the $\tau^+ \to \pi^+ \pi^- \pi^+ (\pi^0) \neutb$ decay.\footnote{Charge conjugated decay modes are implied throughout the document.}
Since the neutrinos and the $\pi^0$ are not detected, the visible final state consists of five pions plus a kaon.
It is experimentally convenient to measure:
\begin{equation}
	\mathcal{R}_{had} ( D^* ) = \frac{ \mathcal{B}\left( B^{0}\to D^{*-} \tau^{+} \nu_{\tau} \right) }{ \mathcal{B}\left( B^{0}\to D^{*-} \pi^+ \pi^- \pi^+ \right) },
\label{eq:RDstar_had}
\end{equation}
because most of the systematic uncertainties will cancel out in the efficiency ratio, since signal and normalization have the same final state.
Once $\mathcal{R}_{had} ( D^* )$ is measured, it will be multiplied by externally measured branching fractions to obtain $\mathcal{R} ( D^* )$:
\begin{equation}
	\mathcal{R} (D^*) = \mathcal{R}_{had} ( D^* ) \times \frac{\mathcal{B}\left( B^{0}\to D^{*-} \pi^+ \pi^- \pi^+ \right) }{ \mathcal{B}\left( B^{0}\to D^{*-} \mu^{+} \nu_{\mu} \right)}.
\label{eq:RDstar_final}
\end{equation}

%\subsection{Background}

The most dominant background consists of inclusive decays of b-hadrons to $D^* 3\pi X$, where the three pions come promptly from the b-hadron decay vertex.
Since the $\tau$ decay vertex is reconstructed with good resolution, it is possible to suppress this kind of background requiring the $\tau$ vertex to be downstream, along the beam direction, with respect to the $B$ vertex with a $4\sigma$ significance.
This selection, applied along with other topological cuts, suppresses the dominant background by three orders of magnitude.

The background surviving the first selection is mainly due to double-charmed $B$ decays, since their topology is very similar to the signal one.
This kind of background is dominated by $B^0 \to D^{*-} D^{+}_s (\to \pi^+ \pi^- \pi^+ X)$ decay, whose branching ratio is 4 times larger than the signal.
In order to discriminate this background from signal, a set of variables is used; they can be grouped in three categories: variables computed with two partial reconstruction techniques, one in signal hypothesis and the other in background hypothesis; isolation variables (i.e. how much the signal tracks are isolated from the other tracks and neutral energy deposits in the event); variables related to the $3\pi$ system dynamics.
These variables are used as input to train a Boosted Decision Tree (BDT).

The partial reconstruction in signal hypothesis allows to compute the squared $B - D^*$ transferred momentum $q^2$ and the $\tau$ decay time with a sufficiently good resolution to maintain separation between signal and background.

%\subsection{Fit Strategy}

Three-dimensional shapes of $q^2$, $\tau$ decay time and BDT output are extracted from simulated and data-driven control samples which represent the various contributions in data.
In order to extract the signal yield, the three-dimensional shapes are used to perform an extended maximum-likelihood template fit on data in high-BDT region ($BDT>-0.075$).
The various templates used in the fit can be grouped in 5 categories: signal (both $\tau^+ \to \pi^+ \pi^- \pi^+ \neutb$ and $\tau^+ \to \pi^+ \pi^- \pi^+ \pi^0 \neutb$), excited $D^*$ contributions (i.e. $B^0 \to D^{**} \tau^+ \nu_\tau$), double-charmed components, $B^0 \to D^{*-} \pi^+ \pi^- \pi^+ X$ background and combinatorial background.

Since the relative fractions of the various $D^+_s \to \pi^+ \pi^- \pi^+ X$ decays are currently not well known, they are measured in the low-BDT region, which is enriched in such decays and where the signal is absent.
Four different templates in $\min[m(\pi^+ \pi^-)]$, $\max[m(\pi^+ \pi^-)]$, $m(\pi^+ \pi^+)$ and $m(3\pi)$ are built, corresponding to: events where at least one pion comes from an $\eta'$ resonance, events where at least one pion comes from an $\eta$ resonance but none of them originates from an $\eta'$, events where the pions come from a resonance which is not $\eta'$ nor $\eta$ and events where the pions do not originate from a resonance.
A template fit is performed, and the resulting relative fractions from the low-BDT region are then used to constrain the $D^+_s$ decay model in the high-BDT region.

%\subsection{Normalization}

To select normalization events, the $\tau$ vertex requirement is reversed, i.e. the $\tau$ vertex is required to be upstream with respect to the $D^0$ vertex with a $4\sigma$ significance.
The normalization yield is obtained by fitting the $D^* 3\pi$ invariant mass distribution (see Figure~\ref{fig:ctrl_samples}) in the $B$ region.

\section{Control samples}

\begin{figure}
\begin{center}
	\includegraphics[width=0.45\linewidth]{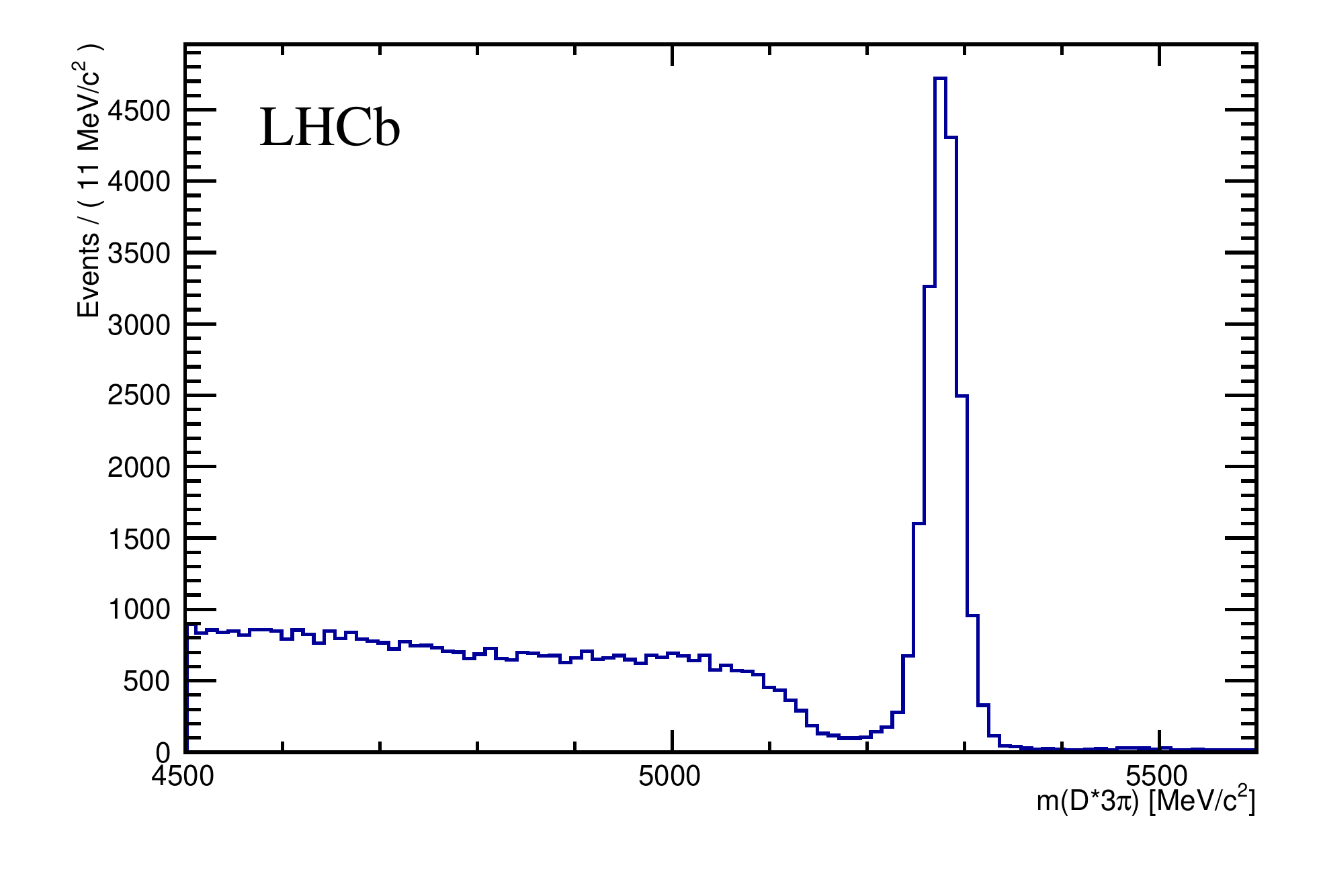}
	\includegraphics[width=0.45\linewidth]{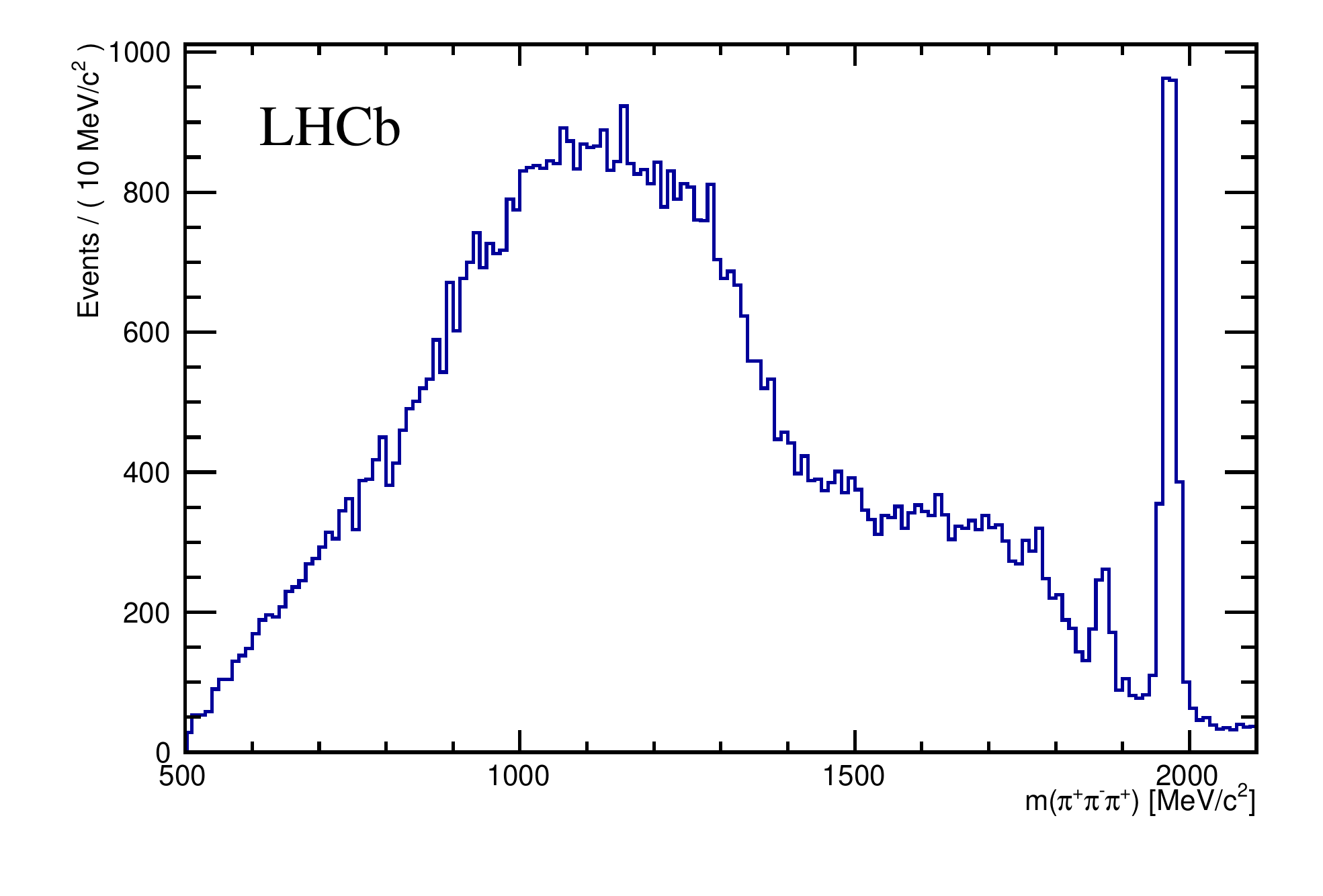}
	\includegraphics[width=0.45\linewidth]{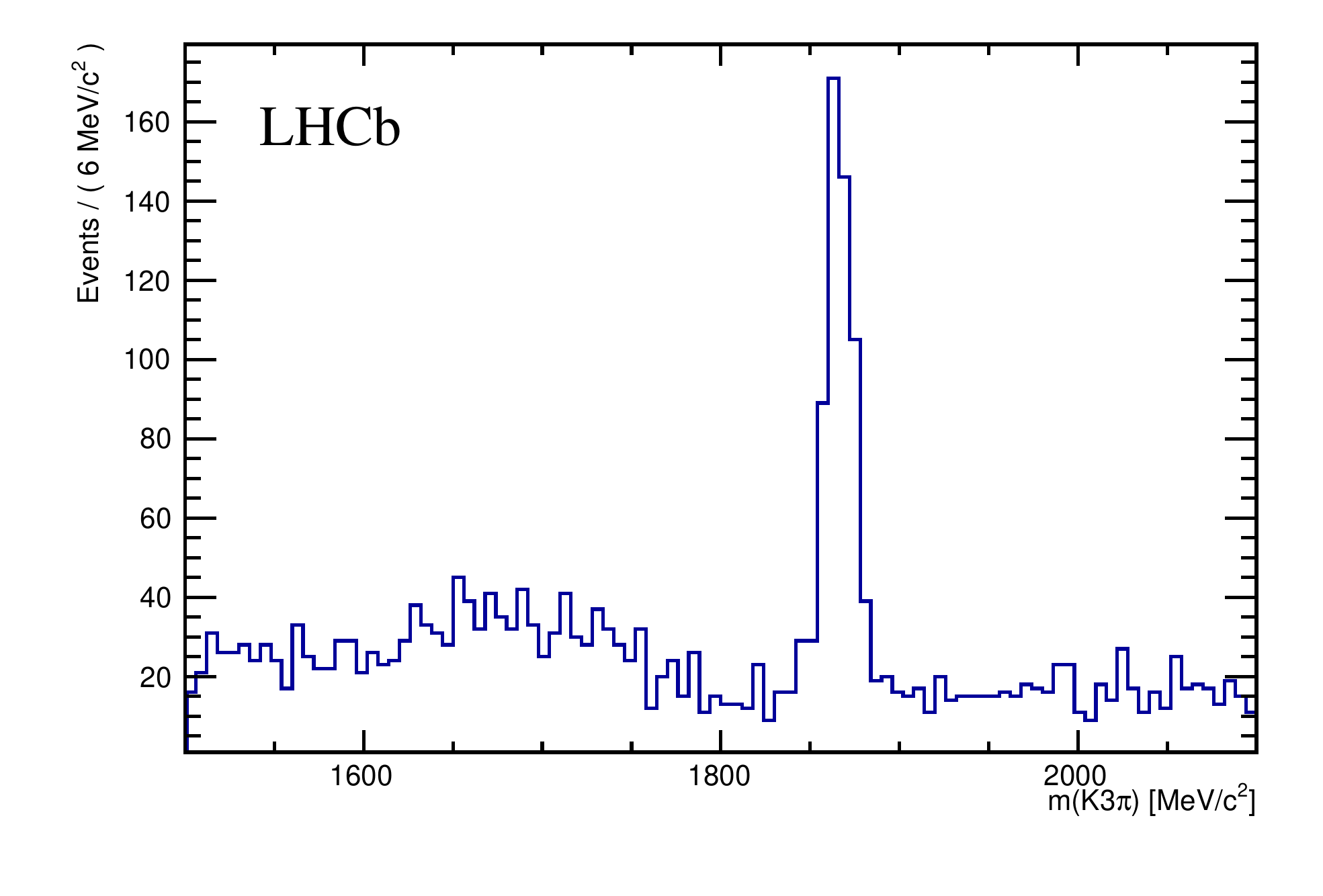}
	\includegraphics[width=0.45\linewidth]{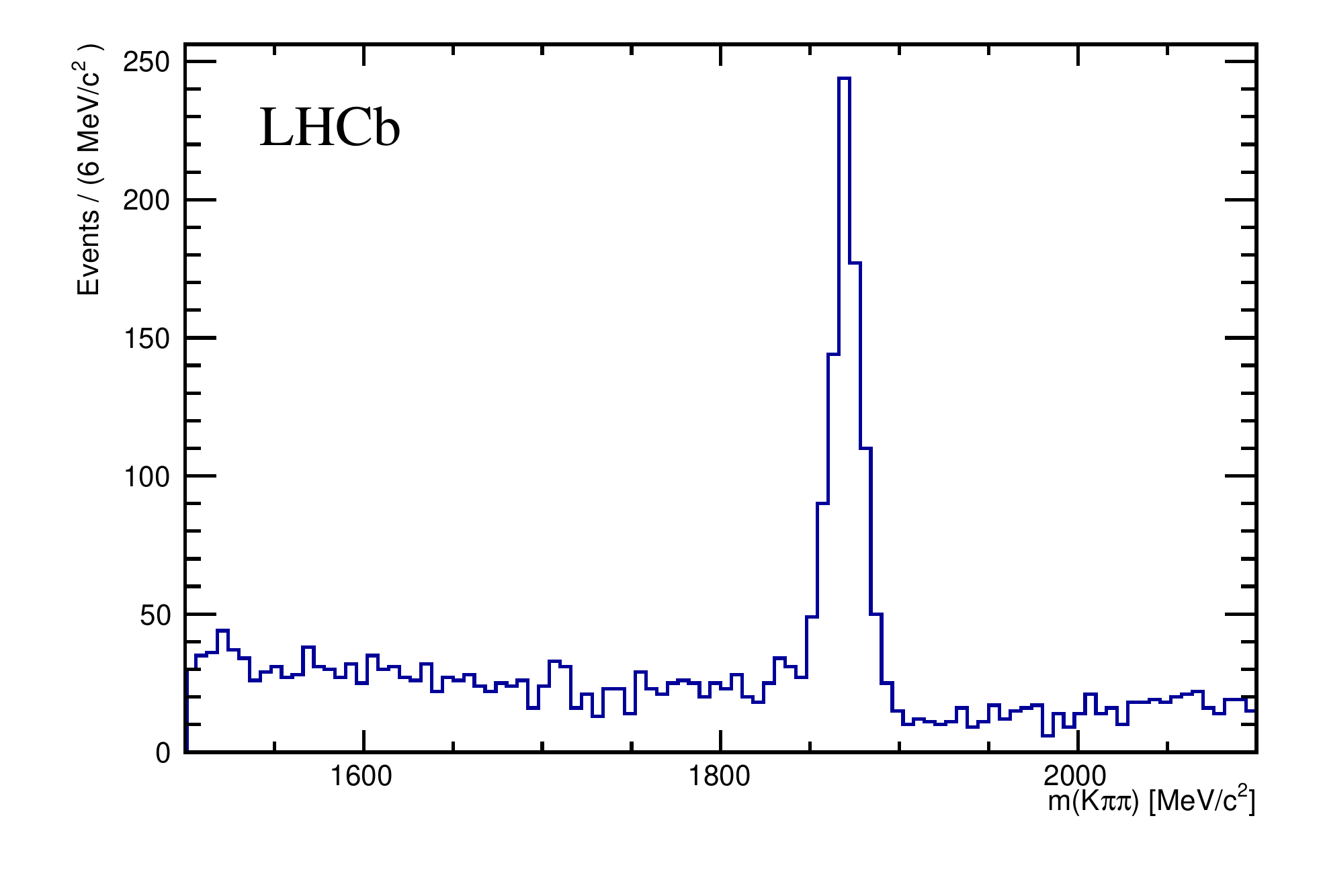}
\end{center}
\caption{
(top left) $D^* 3\pi$ invariant mass distribution for normalization events.
(top right) $\pi^+ \pi^- \pi^+$ invariant mass distribution; the peak in the $D^+_s$ region is used to extract the $B \to D^* D^+_s X$ control sample.
(bottom left) $K^- \pi^+ \pi^- \pi^+$ invariant mass distribution; the peak in the $D^0$ region is used to obtain the $B \to D^* D^0 X$ control sample.
(bottom right) $\pi^+ \pi^- \pi^+$ invariant mass distribution with kaon hypothesis on the $\pi^-$; the peak in the $D^+$ region is needed to build the $B \to D^* D^+ X$ control sample.
}
\label{fig:ctrl_samples}
\end{figure}

In order to validate the simulated samples, three control samples extracted from data are used (see Figure~\ref{fig:ctrl_samples}):
\begin{itemize}
	\item $B \to D^* D^+_s X$ sample, obtained by selecting events in the exclusive $D^+_s \to \pi^+ \pi^- \pi^+$ peak in the $3\pi$ invariant mass distribution.
	\item $B \to D^* D^0 X$ sample, selected by requiring a charged kaon around the $3\pi$ vertex and the $K 3\pi$ invariant mass to be compatible with the $D^0$ mass.
	\item $B \to D^* D^+ X$ sample, obtained by requiring kaon particle identification criteria for the $\pi^-$ in the $\pi^+ \pi^- \pi^+$ system, and the $K^- \pi^+ \pi^+$ to be compatible with the $D^+$ mass.
\end{itemize}

\section{Perspectives}

\begin{figure}[!h]
\begin{center}
	\includegraphics[width=0.65\linewidth]{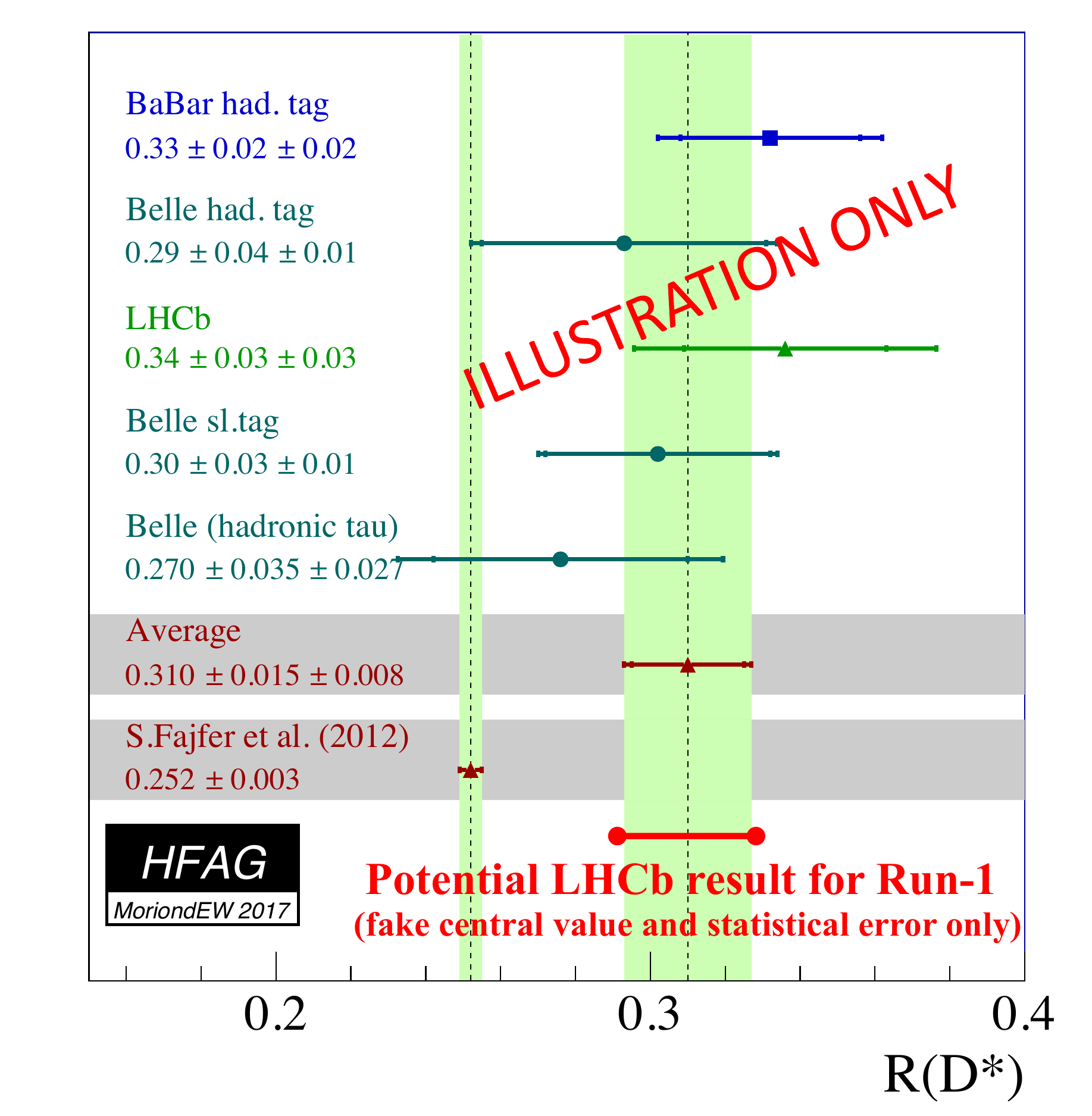}
\end{center}
\caption{Potential LHCb statistical uncertainty with Run 1 data, compared to previous measurements of $\mathcal{R}( D^* )$, the World Average and the SM prediction.}
\label{fig:stat}
\end{figure}

The statistical precision of this measurement is 6.7\%, competitive with the previous LHCb measurement, which used the muonic $\tau$ decay channel, and with the World Average.
This is the smallest statistical uncertainty for a single measurement of $\mathcal{R}(D^*)$.
A comparison with the previous measurements, the World Average and the SM prediction is reported in Figure~\ref{fig:stat}, which is an adapted version of the latest report by HFAG on the status of $\mathcal{R} (D^*)$~\cite{HFAG}.
Here only the statistical uncertainty of this analysis is reported, while the central value is set on the World Average, for illustration purpose.

The largest systematic uncertainties are due to the limited statistics of the simulated samples and to the precision on the knowledge of the external branching ratios $\mathcal{B}\left( B^{0}\to D^{*-} \mu^{+} \nu_{\mu} \right)$ and $\mathcal{B}\left( B^{0}\to D^{*-} \pi^+ \pi^- \pi^+ \right)$.
Another source of systematic uncertainty is due to the knowledge of the various $D^+_s$, $D^+$ and $D^0$ background decay models.

Statistical uncertainty is expected to decrease at least by a factor 2 using Run 2 data, since the $B$ production cross section is higher at 13 TeV and the trigger is more efficient with respect to Run 1.

\end{document}